\documentclass[reprint,aps,pra,superscriptaddress,amsmath,amssymb,longbibliography,nofootinbib]{revtex4-2}
\allowdisplaybreaks
\usepackage{dcolumn}
\usepackage{graphicx}
\usepackage{bm}
\pdfoutput=1
\usepackage[utf8]{inputenc}
\usepackage[english]{babel}
\usepackage[T1]{fontenc}
\usepackage{amsmath}
\usepackage{amssymb}
\usepackage{hyperref}
\usepackage{tikz}
\usepackage{lipsum}

\usepackage{float}

\newcommand{\orcid}[1]{}

\begin{document}

\title{Superradiance in dense atomic samples}

\author{I. M. de Araújo}
\affiliation{Instituto de Física de São Carlos, Universidade de São Paulo, Caixa Postal 369, 13560-970, São Carlos, São Paulo, Brazil}
\thanks{Corresponding author.}
\email{italo.maraujo@ifsc.usp.br}
\orcid{0000-0002-8760-7731}

\author{H. Sanchez}
\affiliation{Instituto de Física de São Carlos, Universidade de São Paulo, Caixa Postal 369, 13560-970, São Carlos, São Paulo, Brazil}
\orcid{0000-0002-4915-4495}

\author{L. F. Alves da Silva}
\affiliation{Instituto de Física de São Carlos, Universidade de São Paulo, Caixa Postal 369, 13560-970, São Carlos, São Paulo, Brazil}
\orcid{0000-0002-8418-4833}

\author{M. H. Y. Moussa}

\affiliation{Instituto de Física de São Carlos, Universidade de São Paulo, Caixa Postal 369, 13560-970, São Carlos, São Paulo, Brazil}
\orcid{0000-0002-3026-0845}

\begin{abstract}
  Here we present an approach to the problem of superradiance in dense atomic
samples, when dipolar interactions arise between atoms. Our treatment consists
of the sequential use of the Holstein-Primakoff and mean-field approximations,
from which we derive master equations for the strong and weak couplings of the
sample with the reservoir. We find, in both cases, that the radiation emission
presents remarkable features, with characteristic emission times much shorter
and intensities much higher than those of Dicke superradiance. In particular,
for strong sample-reservoir coupling, a whole comb of superpulses occurs
within an envelope with the above-mentioned characteristic emission times much
shorter and intensities much higher than those of Dicke superradiance.
\end{abstract}
\maketitle

\section{Introduction}
\label{sec:intro}
Predicted by R. H. Dicke in 1954 \cite{Dicke}, the superradiance is the
collective spontaneous emission of radiation by a moderately dense atomic
sample, spatially confined to a region of dimensions smaller than the
wavelength of the emitted field. This small size sample assumption contrasts
superradiance with the well-known fluorescence, whereby population inversion
of the sample leads to spontaneous emission in a time interval inversely
proportional to the atomic decay rate $\gamma$, with intensity proportional to
the number $N$ of atoms in the sample. Conversely, superradiant emission
occurs with strongly reduced characteristic emission time, $\tau_{c}%
=2/N\gamma$, and enhanced intensity, proportional to $N^{2}$. This collective
phenomenon, occurs when the
entire sample of two-level atoms interacts with a common thermal reservoir. As
a result, in addition to the direct (or diagonal) dissipation channels
---through which the atoms emit directly to the reservoir--- there are
indirect (or non-diagonal) dissipation channels ---through which the atoms
emit indirectly to the reservoir, through their couplings with all the other
atoms--- induce atomic correlations mediated by the reservoir. These
correlations align the individual atomic dipoles into a giant dipole and the
entire sample behaves as a single effective atom. The correlations arising
from non-diagonal dissipation channels lead to other important developments,
such as the construction of decoherence-free \cite{DFS, DFS2, DFS3} or quasi-free
\cite{Castro} subspaces for high-fidelity quantum logical processing, a
central issue in quantum information theory.

From its first experimental observation by N. Skribanowitz \textit{et al}.
\cite{S}, superradiance has been demonstrate in a variety of systems, such as
molecular H aggregates \cite{MHA}, trapped atoms in cavity QED \cite{CTA, CTA2},
cold atoms \cite{CA, CA2}, and Bose-Einstein condensates \cite{BEC, BEC2}. We stress the
observation of two-photon \cite{Brewer} and single-photon superradiance
\cite{SPS, SPS2, SPS3}, with potential applications in the control of spontaneous emission
and ultrafast readout. The phenomenon has always played a prominent role in
the field of atomic optics, advancing many interesting effects and
applications as the superradiant laser \cite{SL,Rodrigo}, phonon superradiance
\cite{Phonon}, and superradiance lattice \cite{Scully}.

In addition to the semiclassical treatment through the Maxwell-Bloch equations
\cite{Haroche}, superradiance is basically approached through the quantum
master equation \cite{Agarwal1}, and from this, we observe the possibility of
deriving a non-linear single-particle mean-field Hamiltonian \cite{Salomon, Salomon2}.
As might be expected, only the non-diagonal dissipation channels contribute to
the derivation of this mean-field Hamiltonian.

More recently, Higgins \textit{et al.} \cite{Higgins} proposed the reciprocal
process to superradiance, the collective superabsorption of light, and in Ref.
\cite{Rodrigo} the coherent many body Rabi-oscillations was achieved through
the interplay between superradiance and superabsorption. To this end a
moderately dense atomic sample was confined within a high-finesse cavity and a
non-linear mean-field Hamiltonian was derived for the interaction between the
sample and the cavity field. When the sample is prepared in a superradiant
state and the cavity field in the vacuum, the superradiant pulse emitted by
the sample is superabsorbed by the resonant cavity field as a consequence to
Rabi coupling $g$ enhanced by the factor $\sqrt{N}$. This enhanced coupling
was previously derived through a semiclassical approach \cite{HM, HM2} and
experimentally confirmed in what is called the ringing regime of superradiance
\cite{Kaluzny}. This enhanced many body Rabi coupling can be used to speed up
logical operations in quantum comunication and computation, with the typical
time of operations going from $1/g$ to $1/\sqrt{N}g$, a dramatic decrease for
mesoscopic samples and protection against decoherence. The mean-field
Hamiltonian derived in \cite{Rodrigo} was subsequently used to achieve the
coherent deflection of an atomic sample, advancing a protocol for the
preparation of positional mesoscopic superpositions \cite{Luis}. From the
prospectus outlined above, we verify that superradiance is thus leaving the
realm of theoretical ideas to become an important tool in experiments on
many-body quantum optics.

Moving from atomic optics to biophysics, a model for neural networks has
recently been proposed in which the neuron is treated as a spin-boson system
and the network as coupled spin-boson units \cite{Hugo}. The aim of the model
is to approach seizure, and it was found that this process can be described,
in close parallel with atomic optics, as a fluorescence- to superradiance-like
phase transition. For the derivation of the master equation describing the
state of the seizure focus, the Holstein-Primakoff \cite{HP} and mean-field
\cite{Salomon} approximations were adopted, and we shall tackle superradiance
in dense atomic samples based on the same procedure. The difference between
the neural network and the sample of interacting atoms is basically in the
quantum tunneling between the states, present in the spin-boson system and
absent in the two-level atoms.

Our goal here, therefore, is the characterization of the coherent pulse
emitted by dense atomic samples, in particular, its emission time and
intensity. We verify the emergence of distinctive features of the radiation
emission from the dipole-dipole coupling between atoms, in particular in the
strong sample-reservoir coupling regime. These features have the potential to
motivate new and varied studies of the collective emission in dense samples. 

The superradiance in the presence of dipolar interaction between atoms has
been little discussed previously. In Ref. \cite{Han}, considering an array of
Rydberg atoms in a dissipative microwave cavity, the authors demonstrate that
the steady state of the system exhibts a Rydberg-interaction-enhanced
superradiance. In Ref. \cite{Mok} it was demonstrated that Dicke superradiance
requires interactions beyond nearest neighbors, while a method for approaching
cooperative radiation emission in many-body systems was presented in Ref.
\cite{Yelin}.

\section{A Dense Atomic Sample Strongly Coupled to the Reservoir}

We first observe that the strong coupling between atomic samples and
reservoirs must be achieved through reservoir engineering techniques
\cite{RE,RE2, RE3, RE4, RE5, RE6, RE7, RE8}, subjecting the atomic sample to interact with a strongly
dissipative auxiliary system, such as a field-mode in the bad-cavity limit,
whose degrees of freedom are subsequently eliminated. Without resorting to
these techniques, the sample will be weakly coupled to the reservoir. However,
it is important to address the case of strong coupling since it results in
very distinctive characteristics in the collective emission of radiation. Our
Hamiltonian for modelling a dense sample of $N$ two-level atoms strongly
coupled to the reservoir is giving by ($\hbar=1$)
\begin{align}
	H &= \omega_{0}\,S_{z} + \sum_{k}\omega_{k}\, b_{k}^{\dagger} b_{k}
	- g \sum_{\substack{m,n=1\\ m\neq n}}^{N}
	\bigl(\sigma_{+}^{(m)}\sigma_{-}^{(n)}+\sigma_{-}^{(m)}\sigma_{+}^{(n)}\bigr) \notag\\
	&\quad
	+ \sum_{k}\lambda_{k}\,(S_{+}+S_{-})\,(b_{k}+b_{k}^{\dagger}) .
	\label{1}
\end{align}
where $\omega_{0}$ is the transition frequency of the atoms described by the
pseudo-spin operator $S_{z}= \sum_{m=1}^{N}\sigma_{z}^{m}/2$, with $\sigma_{\mu}$ 
being the Pauli spin operators with $\mu=x,y,z$ and $\sigma_{\pm}=\left(\sigma_{x}\pm i\sigma_{y}\right)/2$.
The multimodal reservoir of frequencies $\omega_{k}$ is described by the set
of bosonic creation and annihilation operators $\left\{  b_{k}^{\dagger
}\right\}  $ and $\left\{  b_{k}\right\}  $, respectively. The third term
describes the dipole-dipole interaction $g$ between the atoms, assuming for
simplicity a symmetric coupling where all atoms interact pairwise. The
negative sign favors the alignment of the electric dipoles, as expected,
similar to ferromagnetism. The coupling $\lambda_{k}$ between the sample and
the reservoir, beyond the RWA approximation, is described by the last term of
the Hamiltonian.

Reasoning by analogy with Ref. \cite{Mickel}, where a general treatment for
bosonic dissipative networks is presented, for the derivation of the master
equation of the sample we must first diagonalize the interaction between the
atoms, leaving aside their interaction with the reservoir. For this, we resort
to the collective Holstein-Primakoff transformation, a canonical mapping of
collective pseudo-spin operators onto global bosonic creation and annihilation
operators $A^{\dagger}$ and $A$. Under the assumption $N\gg\langle A^{\dagger
}A\rangle$, it follows that%

\begin{subequations}
\begin{align}
S_{+}  &  =\sqrt{N-A^{\dagger}A}A\approx\sqrt{N}A,\label{2a}\\
S_{-}  &  =A^{\dagger}\sqrt{N-A^{\dagger}A}A^{\dagger}\approx\sqrt
{N}A^{\dagger},\label{2b}\\
S_{z}  &  =\frac{N}{2}-A^{\dagger}A, \label{2c}%
\end{align}
leading the Hamiltonian (\ref{1}), to the simplified form
\end{subequations}
\begin{equation}
\tilde{H}=\Omega A^{\dagger}A+\sum_{k}\omega_{k}b_{k}^{\dagger}b_{k}+\sum
_{k}\sqrt{N}\lambda_{k}\left(  A+A^{\dagger}\right)  \left(  b_{k}%
+b_{k}^{\dagger}\right)  , \label{3}%
\end{equation}
where an effective resonator interacts strongly with the reservoir. The
Holstein-Primakoff transformation thus introduces an effective frequency
$\Omega=(1+\alpha)\omega_{0}$, along with the collective interaction parameter
$\alpha=2gN/\omega_{0}$, which accounts for a collective interatomic coupling. It
is worth noting that we could have expected that the sample of interacting
atoms would in fact reduce to an effective resonator; after all, when we
consider that all the atoms interact with each other, we approach what we can
describe as a Bose-Einstein condensate.

Using the Caldeira-Leggett developments \cite{CL, CL2} on the Feynman-Vernon
influence-functional method \cite{FV}, we automatically derive from
Hamiltonian (\ref{3}) the master equation governing the evolution of the
density operator $\rho_{N}$ for the atomic sample. We obtain%
\begin{align}
	\dot{\rho}_{N}
	&= i\Omega\,[A^{\dagger}A,\rho_{N}]
	- iN j(\Omega)\,[X,\{P,\rho_{N}\}] \notag\\
	&\quad
	- N j(\Omega)\,\coth\!\left(\frac{\Omega}{k_{B}T}\right)
	[X,[X,\rho_{N}]] .
	\label{4}
\end{align}
where $X=\left(  A+A^{\dagger}\right)  /2$, $P=i\left(  A-A^{\dagger}\right)
/2$, $k_{B}$ is the Boltzmann constant, $j(\Omega)$ and $T$ are the spectral
density and temperature of the reservoir.

Next, rewriting the master equation back to the pseudo-spin operators, we
obtain
\begin{equation}
\dot{\rho}_{N}=-i\left[  H_{S},\rho_{N}\right]  +\mathcal{L}\rho_{N}%
\text{,}\label{5}%
\end{equation}
where $H_{S}=\Omega S_{z}$ is the Hamiltonian for the atomic sample and the
Liouvillian is given by
\begin{widetext}
	\begin{equation}
			\label{6}
		\begin{aligned}
			\mathcal{L}\rho_{N}
			&= \sum_{m,n=1}^{N}
			\frac{\delta_{mn}\gamma + (1-\delta_{mn})\Gamma}{2}\,
			\Bigl(
			[\sigma_{+}^{(m)},\,\sigma_{-}^{(n)}\rho_{N}]
			- [\sigma_{-}^{(m)},\,\rho_{N}\sigma_{+}^{(n)}] \\
			&\qquad\qquad\qquad
			- [\sigma_{+}^{(m)},\,\rho_{N}\sigma_{+}^{(n)}]
			+ [\sigma_{-}^{(m)},\,\sigma_{-}^{(n)}\rho_{N}]
			\Bigr).
		\end{aligned}
	\end{equation}
\end{widetext}
accounting for the diagonal ($m=n$) and non-diagonal ($m\neq n$) dissipative
channels. Assuming Ohmic spectral density, we obtain the effective dissipative
factor $\Gamma=\left(  1+\alpha\right)  \gamma=2j(\Omega)$, weighting the
non-diagonal channels, where $\gamma=2j(\omega_{0})$ stands for the atomic
decay rate. Next, we proceed to the mean-field approaximation by computing the
density operator for $p<N$ atoms, i.e., by tracing out the degrees of freedom
of $N-p$ atoms, under the assumption that $%
{\textstyle\sum\nolimits_{r=p+1}^{N}}
\operatorname{Tr}_{p+1,...,N}\sigma_{\pm}^{r}\rho_{N}\approx
(N-p)\operatorname{Tr}_{p+1,...,N}\sigma_{\pm}^{r}\rho_{N}$ for $r>p$. From
this mean-field assumption we derive the $p$-body master equation
\cite{Salomon}
\begin{widetext}
\begin{align}
	\dot{\rho}_{p}
	&= -\,i\,\frac{\Omega}{2}\,\sum_{m=1}^{p}\,[\,\sigma_{z}^{m},\,\rho_{p}\,]
	\notag\\[2pt]
	&\quad
	+ \frac{\Gamma}{2}\,\sum_{\substack{m,n=1\\ (m\neq n)}}^{p}
	\Big(\,[\,\sigma_{+}^{m},\,\sigma_{-}^{n}\rho_{p}\,]
	-[\,\sigma_{-}^{m},\,\rho_{p}\sigma_{+}^{n}\,]
	+[\,\sigma_{-}^{m},\,\sigma_{-}^{n}\rho_{p}\,]\Big)
	\notag\\[2pt]
	&\quad
	+ (N-p)\,\frac{\Gamma}{2}\,\sum_{m=1}^{p}
	\Big(\,[\,\sigma_{+}^{m},\,\mathrm{Tr}_{p+1}\,\sigma_{+}^{p+1}\rho_{p+1}\,]
	-[\,\sigma_{-}^{m},\,\mathrm{Tr}_{p+1}\,\rho_{p+1}\sigma_{+}^{p+1}\,]
	\notag\\[-1pt]
	&\qquad\qquad\qquad\qquad
	-[\,\sigma_{+}^{m},\,\mathrm{Tr}_{p+1}\,\rho_{p+1}\sigma_{+}^{p+1}\,]
	+[\,\sigma_{-}^{m},\,\mathrm{Tr}_{p+1}\,\sigma_{-}^{p+1}\rho_{p+1}\,]\Big).
	\label{R}
\end{align}
\end{widetext}
From the final assumption of uncorrelated two-body states, such that $\rho
_{2}=\rho_{1}\otimes\rho_{1}$, we finaly obtain, for $p=1$, the master
equation for the representative atom of the sample:
\begin{align}
	\dot{\rho}
	&= -\,i\,[\,\mathcal{H},\,\rho\,] \notag\\[2pt]
	&\quad
	+ \frac{\Gamma}{2}\Bigl(
	[\,\sigma_{+},\,\sigma_{-}\rho\,]
	- [\,\sigma_{-},\,\rho\sigma_{+}\,] \notag\\[2pt]
	&\qquad\qquad
	- [\,\sigma_{+},\,\rho\sigma_{+}\,]
	+ [\,\sigma_{-},\,\sigma_{-}\rho\,]
	\Bigr).
	\label{8}
\end{align}
where the non-linear mean-field Hamiltonian becomes
\begin{equation}
\mathcal{H}=\frac{\Omega}{2}\sigma_{z}+(N-1)\frac{\Gamma}{4}\left\langle
\sigma_{y}\right\rangle \sigma_{x}.\label{9}%
\end{equation}

Interested in a short-time phenomenon, much shorter than the effective
relaxation time of the sample, $\Gamma^{-1}$, we can safely disregard the
incoherent effects arising from the superoperator in the master equation
(\ref{8}). Therefore, we are left with solving the Schr\"{o}dinger equation
governed by the time-dependent Hamiltonian $\mathcal{H}$. For this we resort
to the Lewis \& Riesenfeld method \cite{LR}, which relies on choosing an
appropriate dynamical invariant $I(t)$ for $\mathcal{H}(t)$, given by%
\begin{equation}
\frac{d}{dt}I(t)=\frac{\partial}{\partial t}I(t)-i\left[  I(t),\mathcal{H}%
(t)\right]  =0.\label{I}%
\end{equation}
Proceeding along the lines developed in Ref. \cite{Salomon1}, we propose the
invariant
\begin{equation}
I=\langle\sigma_{x}\rangle\sigma_{x}+\langle\sigma_{y}\rangle\sigma
_{y}+\langle\sigma_{z}\rangle\sigma_{z},\label{10}%
\end{equation}
whose expectation value $\langle I\rangle=\langle\sigma_{x}\rangle^{2}%
+\langle\sigma_{y}\rangle^{2}+\langle\sigma_{z}\rangle^{2}=R^{2}$ is a
constant of motion that defines the radius $R$ of the Bloch sphere, where the
polar and azimuthal angles $\theta$ and $\phi$ can be used to parametrize the
expectation values $\langle\sigma_{x}\rangle=R\sin{\theta}\cos{\phi}$,
$\langle\sigma_{y}\rangle=R\sin{\theta}\sin{\phi}$, and $\langle\sigma
_{z}\rangle=R\cos{\theta}$. The solution of the Schr\"{o}dinger equation
$i\partial_{t}|\psi(t)\rangle=\mathcal{H}|\psi(t)\rangle$ in terms of the
dynamical invariant is given by
\begin{equation}
|\psi(t)\rangle=e^{i\Phi(t)}%
\begin{pmatrix}
\cos{\left[  \theta(t)/2\right]  }\\
e^{i\phi(t)}\sin{\left[  \theta(t)/2\right]  }%
\end{pmatrix}
,\label{11}%
\end{equation}
where $\Phi(t)=-\omega_{0}t/2$ is the so-called Lewis \& Riesenfeld phase, and
from Eq. (\ref{I}) we verify that the Bloch angles satisfy the coupled
equations
\begin{subequations}
\label{12}%
\begin{align}
\dot{\theta} &  =\left(  N-1\right)  \frac{\Gamma}{2}\sin{\theta}\sin^{2}%
{\phi},\label{12a}\\
\dot{\phi} &  =\Omega-\left(  N-1\right)  \frac{\Gamma}{4}\cos{\theta}%
\sin2{\phi}.\label{12b}%
\end{align}

The energy of the representative atom, fixing $R=1$, is thus given by
\end{subequations}
\begin{equation}
\varepsilon(t)=\frac{\Omega}{2}\left\langle \psi(t)\right\vert \sigma
_{z}\left\vert \psi(t)\right\rangle =\frac{\Omega}{2}\cos{\theta
(t)},\label{13}%
\end{equation}
from which we compute the intensity of the field radiated by the sample%
\begin{equation}
\mathcal{I}(t)=-N\dot{\varepsilon}(t)=\frac{1}{4}N\left(  N-1\right)
\Gamma\Omega\sin^{2}{\theta(t)}\sin^{2}{\phi(t)}.\label{14}%
\end{equation}
showing the typical quadratic dependence on $N$, a hallmark of Dicke
superradiance, apart from the dependence on the effective atomic coupling
$\alpha$ and the sinusoidal functions on $\theta$ and $\phi$.

\section{Dense Atomic Sample Weakly Coupled to the Reservoir}

In the weak coupling regime the Hamiltonian (\ref{1}) takes the form%
\begin{align}
	H &= \omega_{0}\,S_{z} + \sum_{k}\omega_{k}\,b_{k}^{\dagger}b_{k}
	- g \sum_{\substack{m,n=1\\ m\neq n}}^{N}
	\bigl(\sigma_{+}^{(m)}\sigma_{-}^{(n)}+\sigma_{-}^{(m)}\sigma_{+}^{(n)}\bigr) \notag\\
	&\quad + \sum_{k}\lambda_{k}\,\bigl(S_{-}b_{k}+S_{+}b_{k}^{\dagger}\bigr).
	\label{15}
\end{align}
while Hamiltonian (\ref{3}), following from the Holstein-Primakoff
transformation, becomes%
\begin{equation}
\tilde{H}=\Omega A^{\dagger}A+\sum_{k}\omega_{k}b_{k}^{\dagger}b_{k}+\sum
_{k}\lambda_{k}\sqrt{N}\left(  Ab_{k}^{\dagger}+A^{\dagger}b_{k}\right)
,\label{16}%
\end{equation}
displaying the RWA approximation for the sample-reservoir coupling. Here we
observe that the counter-rotating terms of the sample-reservoir coupling in
the last term of Hamiltonian (\ref{15}) was assumed exactly for the derivation
of the rotating terms in the transformed Hamiltonain (\ref{16}). This
procedure guarantees the derivation of the equations from which we correctly
recover the Dicke superradiance when disregarding the  \  interaction between
the atoms and their strong couplings with the reservoir.

The master equation preserves the form of Eq. (\ref{5}), with the same
Hamiltonian for the atomic sample, $H_{S}=\Omega S_{z}$, but the Liouvillian
changed to%
\begin{equation}
	\label{17}
	\begin{aligned}
		\mathcal{L}\rho_{N}
		&= \sum_{m,n=1}^{N}\frac{\delta_{mn}\gamma+(1-\delta_{mn})\Gamma}{2}
		\\
		&\quad\times\Bigl(\,[\,\sigma_{+}^{m},\,\rho_{N}\sigma_{-}^{n}\,]
		-\,[\,\sigma_{-}^{m},\,\sigma_{+}^{n}\rho_{N}\,]\,\Bigr).
	\end{aligned}
\end{equation}

From the mean-field approximation, the master equation for the representative
atom is now given by%
\begin{equation}
\dot{\rho}(t)=-i\left[  \mathcal{H},\rho(t)\right]  +\frac{\Gamma}{2}\left(
\left[  \sigma_{{+}},\rho(t)\sigma_{-}\right]  -\left[  \sigma_{-},\sigma
_{+}\rho(t)\right]  \right)  ,\label{18}%
\end{equation}
where%

\begin{equation}
\mathcal{H}=\frac{\Omega}{2}\sigma_{z}+(N-1)\frac{\Gamma}{4}\left(
\left\langle \sigma_{x}\right\rangle \sigma_{y}-\left\langle \sigma
_{y}\right\rangle \sigma_{x}\right)  . \label{19}%
\end{equation}

By disregarding the dipole-dipole interaction between the atoms ($g=\alpha
=0$), as occurs in the case of a moderately dense sample, we recover exactly
the master equation for the Dicke's superradiance, with $\Omega$ reducing to
$\omega_{0}$. Again, neglecting the incoherent effects in the master equation
(\ref{18}), for short-time phenomenon, we are left with the Schrodinger
equation $i\partial_{t}|\psi(t)\rangle=\mathcal{H}|\psi(t)\rangle$ whose
solution from the Lewis \& Riesenfeld theorem is given by Eq. (\ref{11}) with
$\Phi(t)=-\Omega t/2$ and
\begin{subequations}
\label{20}%
\begin{align}
\dot{\theta} &  =(N-1)\frac{\Gamma}{2}\sin{\theta},\label{20a}\\
\dot{\phi} &  =2\Omega.\label{20b}%
\end{align}
Differently from Eqs. (\ref{12}), the solutions for the coupled parameters
$\theta$ and $\phi$ follow straightforwardly, given by
\end{subequations}
\begin{subequations}
\label{S}%
\begin{align}
\sin\theta &  =\operatorname{sech}\frac{\left(  t-t_{0}\right)  }{\tau_{c}%
},\label{Sa}\\
\phi(t) &  =\phi_{0}+\Omega t\text{.}\label{Sb}%
\end{align}
The energy of the representative atom is now given by
\end{subequations}
\begin{equation}
\varepsilon(t)=-\frac{\Omega}{2}\tanh\left(  \frac{t-t_{0}}{\tau_{c}}\right)
,\label{21}%
\end{equation}
where $t_{0}=\tau_{c}\ln{N}$ and $\tau_{c}=2/(1+\alpha)N\gamma$
\cite{Salomon1} are the delay and characteristic times, respectively. In the
case where the dense atomic sample is strongly coupled to the reservoir, the
delay and characteristic times cannot be computed analitically. Finally, for a
dense atomic sample weakly coupled to the reservoir, the emitted intensity is
given by
\begin{equation}
\mathcal{I}(t)=\frac{1}{4}N^{2}\Omega\Gamma\operatorname{sech}^{2}\left(
\frac{t-t_{0}}{\tau_{c}}\right)  ,\label{22}%
\end{equation}
showing that it scales as $N^{2}$, but increased by the equally quadratic
factor $\left(  1+\alpha\right)  ^{2}$ coming from the product $\Omega\Gamma$.

Therefore, when considering a dense atomic sample interacting weakly with the
reservoir, the effective atomic coupling $\alpha$ shortens both the delay and
the characteristic times. In addition, the intensity of the emitted radiation
is increased by the factor $\left(  1+\alpha\right)  ^{2}$, which can be
significantly large, so that the radiation emission of dense samples weakly
coupled to the reservoir markedly accentuates the distinctives features of the
Dicke superradiance. The most interesting and new features, however, arise
with the superradiance of dense samples strongly coupled to the reservoir as
discussed bellow.

\subsection{Dicke Superradiance}

The Dicke superradiance of a moderately dense sample interacting weakly with
the reservoir is immediately retrieved from the master equation (\ref{18}),
when considering $g=\alpha=0$. We automatically recover the well-known
characteristic time $\tau_{c}=2/N\gamma$, plus the energy and intensity of the
emitted radiation, in agreement with the values derived in Ref. \cite{Salomon}. This automatic and correct derivation, from the developments in Sections II
and III, of the intensity $\mathcal{I}(t)=\left(  N/2\right)  ^{2}\gamma
\omega_{0}\operatorname{sech}^{2}\left[  \left(  t-t_{0}\right)  /\tau
_{c}\right],$ for Dicke superradiance, is evidence that our procedures for
approaching dense sample superradiance seems quite reasonable.

\section{Radiation Emission for Dense Atomic Samples Strongly Coupled to the
Reservoir}

In order to characterize the superradiant emission of dense samples strongly
coupled to the reservoir, we observe that a reasonable parameter to specify an
effective system-reservoir coupling strength is the ratio $N\gamma/\omega_{0}%
$, assuming, roughly, that the strong coupling follows from \( \frac{N\gamma}{\omega_0} \gtrsim 10^{-2} \); below this value, we must use the the master
equation (\ref{18}) instead of (\ref{8}). 

\bigskip

\bigskip

In Figs. $1$ to $5$ we fix the number of atoms $N$, as well as the frequencies
$\omega_{0}$ and $g$ in units of $\gamma$, to plot the dimensionless scaled
$(a)$ mean energy $\varepsilon(t)/\omega_{0}$ and $(b)$ intensity
$\mathcal{I}(t)/\gamma\omega_{0}$ against $\gamma t$. Starting with $N=10^{4}%
$, $\omega_{0}=10^{6}\gamma$ and $g=10^{2}\gamma$, such that $N\gamma
/\omega_{0}=10^{-2}$ and $\alpha=2$, it is remarkable to observe in Fig.
$1(b)$ that instead of a single superradiant pulse, we now have a comb of
superpulses, the envelope defining a scaled characteristic emission time
\begin{equation}
\gamma\tau_{c}\sim\frac{1}{\left(  1+\alpha\right)  N},\label{24}%
\end{equation}
in agreement with the expression derived for dense samples weakly coupled to
the reservoir. Therefore, in what would be the equivalent of the
characteristic time of Dicke's pulse, we now have a comb of short-duration
superpulses, with characteristic time%
\begin{equation}
\gamma\tau_{1}\sim\frac{\gamma}{(1+\alpha)\omega_{0}}.\label{25}%
\end{equation}
The comb of superpulses results from the oscillatory decay of the mean energy
indicated in Fig. $1(a)$.

\begin{figure*}[t]
	\centering
	\begin{minipage}{0.48\linewidth}
		\centering
		\includegraphics[width=\linewidth]{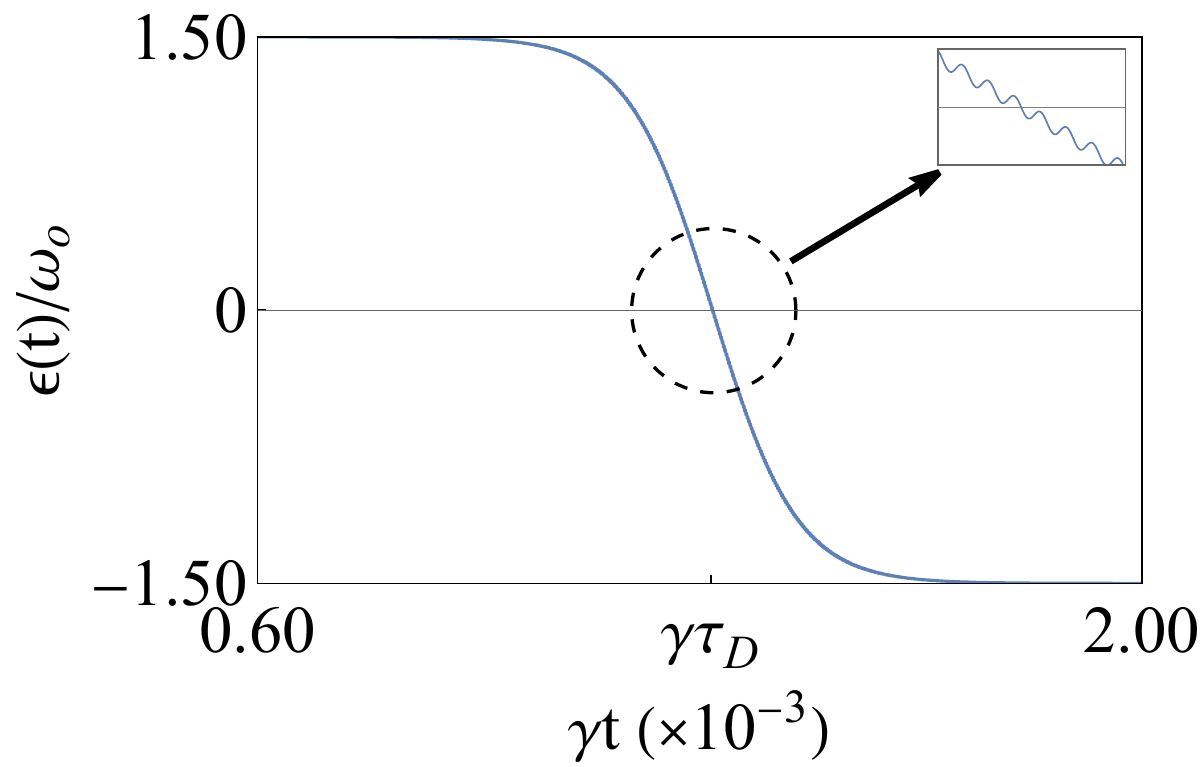}
		\\[-2pt](a)
	\end{minipage}\hfill
	\begin{minipage}{0.48\linewidth}
		\centering
		\includegraphics[width=\linewidth]{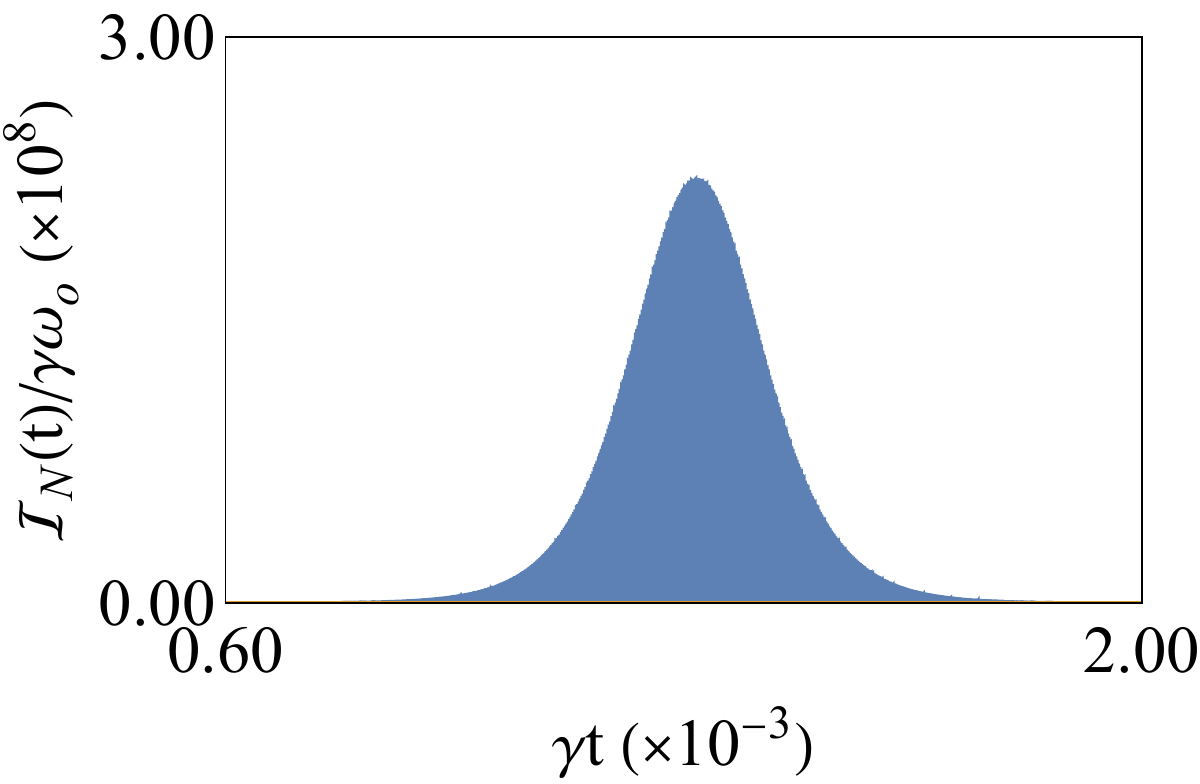}
		\\[-2pt](b)
	\end{minipage}
	\caption{Plot of the scaled (a) mean energy $\varepsilon(t)/\omega_{0}$ and
		(b) intensity $\mathcal{I}(t)/(\gamma\omega_{0})$ against $\gamma t$, for
		$N=10^{4}$, $\omega_{0}=10^{6}\gamma$, and $g=10^{2}\gamma$.}
	\label{fig:energy_intensity}
\end{figure*}

For a rough estimate of the number of superpulses at half-height of the
envelope, we have
\begin{equation}
\frac{\tau_{c}}{\tau_{1}}\sim\frac{\omega_{0}}{N\gamma}, \label{26}%
\end{equation}
the inverse of the effective system-reservoir coupling strength, yielding, for
Fig. $1$, the result $\tau_{1}/\tau_{c}\approx10^{2}$. The scaled intensity of
these superpulses reaches a value proportional to
\begin{equation}
\frac{\mathcal{I}(t)}{\gamma\omega_{0}}\sim\frac{\left[  \left(
1+\alpha\right)  N\right]  ^{2}}{4}, \label{27}%
\end{equation}
while the delay time, indicated in Fig. $1(a)$, is given by
\begin{equation}
\tau_{D}\sim\tau_{c}\ln N=\frac{\ln N}{\left(  1+\alpha\right)  N}. \label{28}%
\end{equation}
both expressions, (\ref{27}) and (\ref{28}), are also in agreement with the
case of dense samples weakly coupled to the reservoir, which seems to indicate
that the dipolar coupling between the atoms, which defines $\alpha$, is more
relevant to the radiation emission process than the coupling strength with the
reservoir. However, we should verify bellow that the coupling strength also
plays a relevant role in the process.

Therefore, as in the case of weak coupling with the reservoir, depending on
the magnitude of alpha, the superpulses resulting from the strong coupling
between the sample and the reservoir, can exhibit considerably shorter
characteristic times and considerably higher intensities than those of Dicke superradiance.

In Fig. $2$ we consider the same parameters as in Fig. $1$, with the exception
of $\omega_{0}=10^{5}\gamma$, resulting in higher values for both
$N\gamma/\omega_{0}=10^{-1}$ and $\alpha=20$. Therefore, the characteristic
times of the superpulses (internal to the envelope), is around the same as
that in Fig. $1$. The decrease in the ration $\omega_{0}/\gamma$ ---the
increased of the decay rate $\gamma$ relative to $\omega_{0}$--- results in a
decrease in the number of superpulses in the envelope, as can be seen in Fig.
$2(b)$. Here, the number of superpulses at half-height of the envelope is
$\tau_{c}/\tau_{1}\approx10$, and the intensity of the emitted radiation is
two orders of magnitude greater than that in Fig. $1$ due to the higher value
of $\alpha$. We emphasize that all the proportionality relations derived in
Fig. $1$ are confirmed by Fig. 2 and, roughly, for all the figures that follow.

\begin{figure*}[t]
	\centering
	\begin{minipage}{0.48\linewidth}
		\centering
		\includegraphics[width=\linewidth]{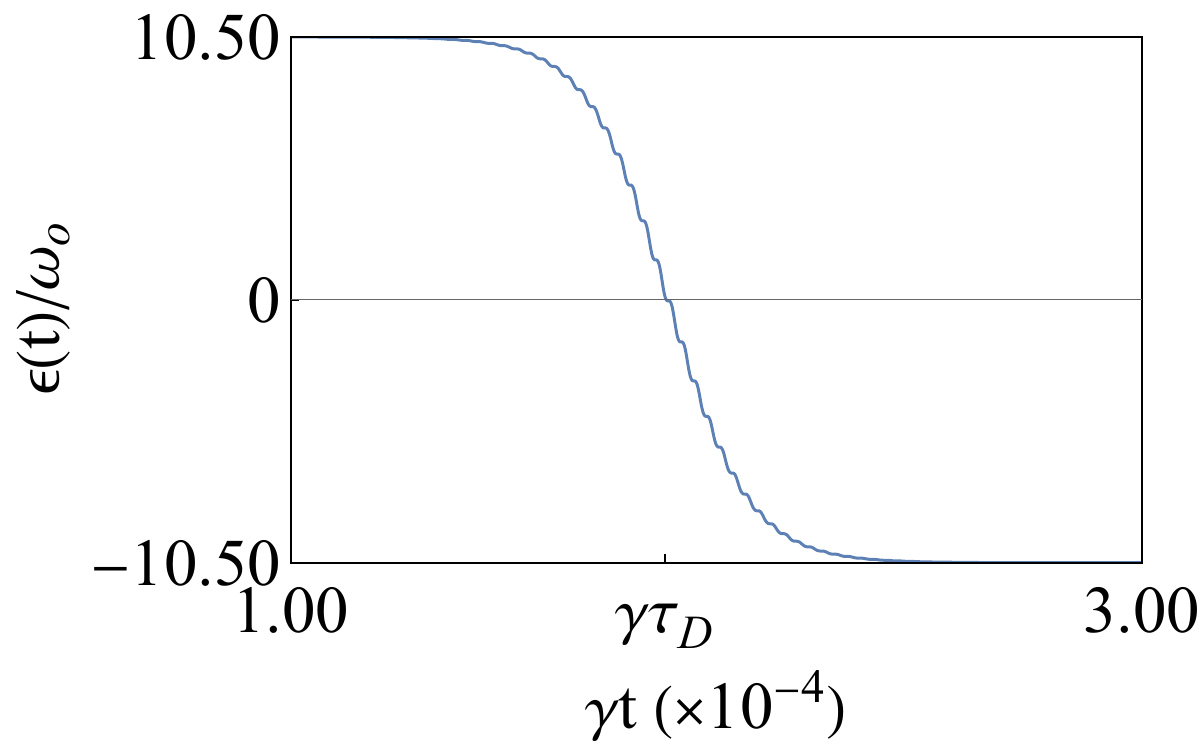}
		\\[-2pt](a)
	\end{minipage}\hfill
	\begin{minipage}{0.48\linewidth}
		\centering
		\includegraphics[width=\linewidth]{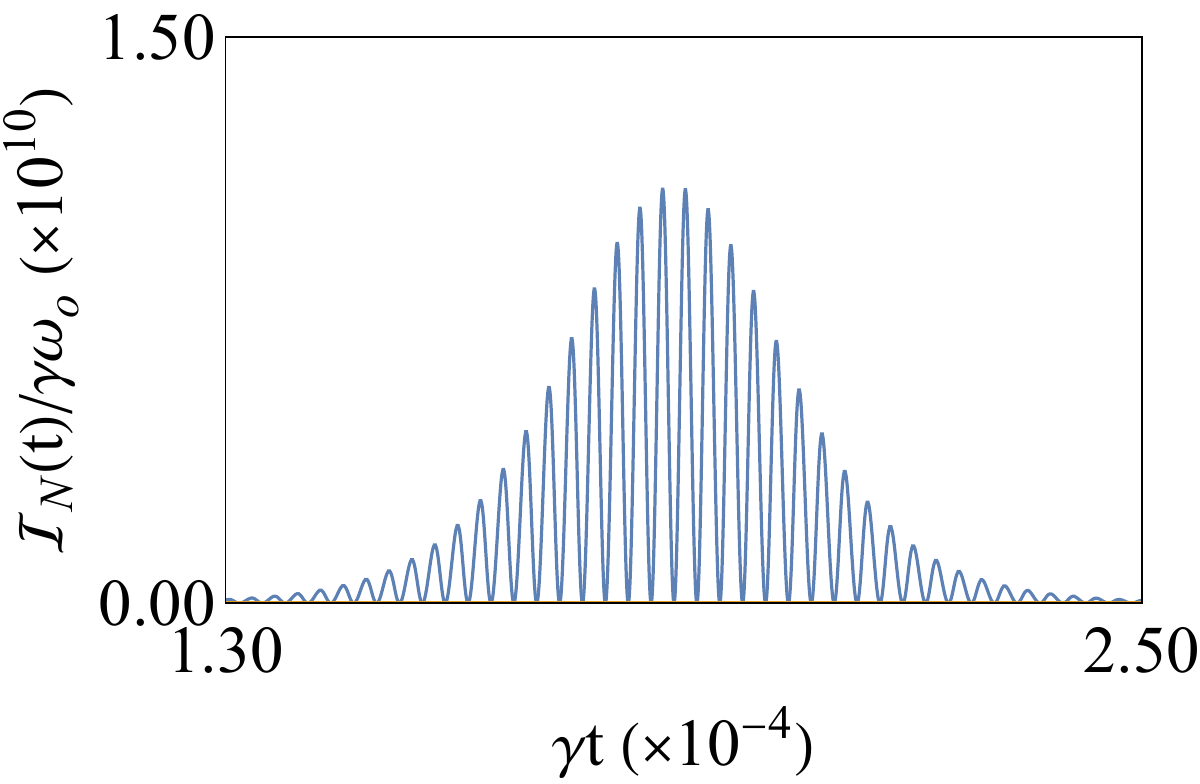}
		\\[-2pt](b)
	\end{minipage}
	\caption{Plot of the scaled (a) mean energy $\varepsilon(t)/\omega_{0}$ and
		(b) intensity $\mathcal{I}(t)/(\gamma\omega_{0})$ against $\gamma t$, for
		$N=10^{4}$, $\omega_{0}=10^{5}\gamma$, and $g=10^{2}\gamma$.}
	\label{fig:energy_intensity_2}
\end{figure*}

Considering the same parameters as in Fig. $1$, with the exception of
$g=10^{3}\gamma$, In Fig. $3$ we keep the ratio $N\gamma/\omega_{0}=10^{-2}$
but increasing $\alpha=20$. Now, both scaled characteristic times, of the
envelope and the superpulses, decrease at the same rate relative to the values
of Fig. $1$ (from $10^{-4}$ and $10^{-6}$ to $10^{-5}$ and $10^{-7}$), thus
keeping the number os superpulses $\tau_{c}/\tau_{1}\approx10^{2}$.

\begin{figure*}[t]
	\centering
	\begin{minipage}{0.48\linewidth}
		\centering
		\includegraphics[width=\linewidth]{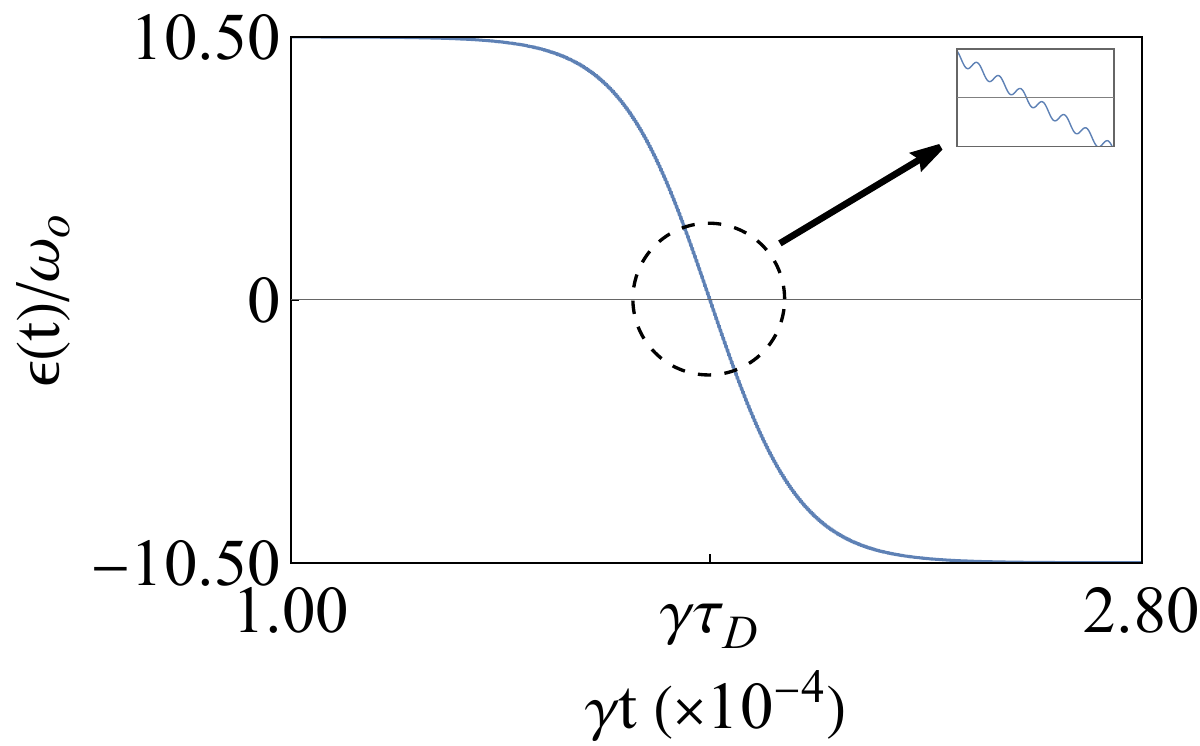}
		\\[-2pt](a)
	\end{minipage}\hfill
	\begin{minipage}{0.48\linewidth}
		\centering
		\includegraphics[width=\linewidth]{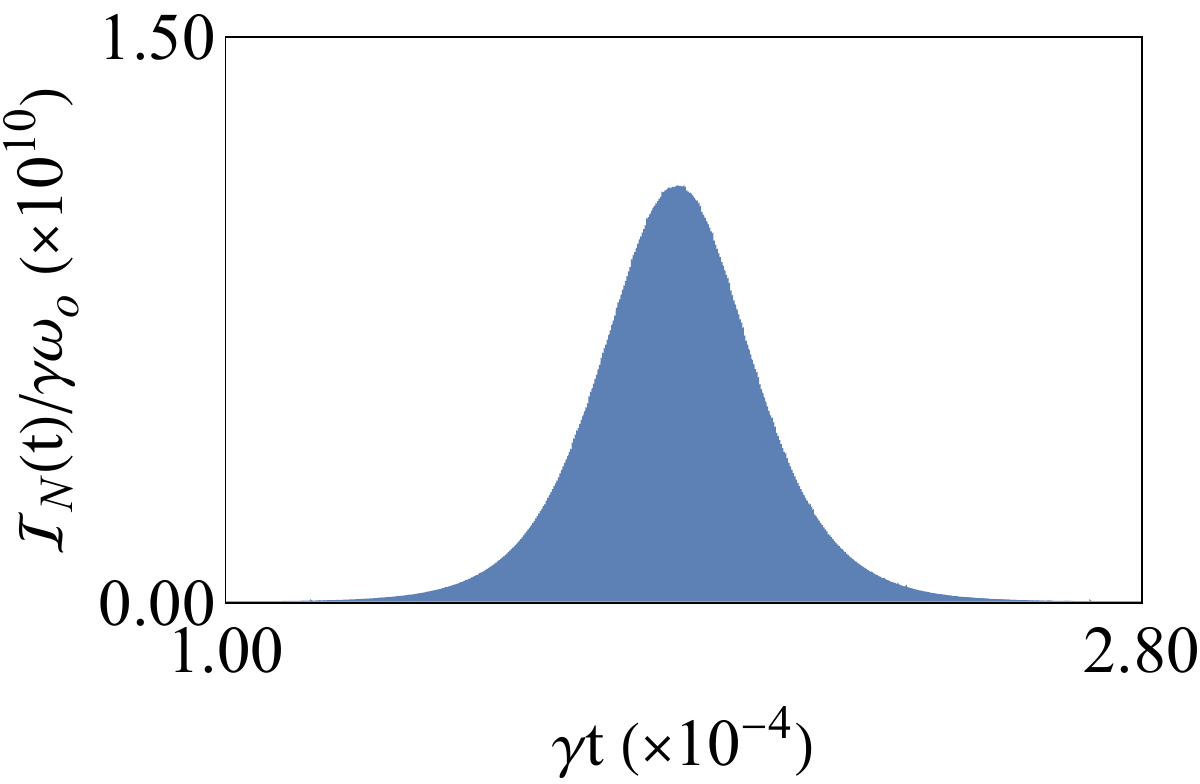}
		\\[-2pt](b)
	\end{minipage}
	\caption{Plot of the scaled (a) mean energy $\varepsilon(t)/\omega_{0}$ and
		(b) intensity $\mathcal{I}(t)/(\gamma\omega_{0})$ against $\gamma t$, for
		$N=10^{4}$, $\omega_{0}=10^{6}\gamma$, and $g=10^{3}\gamma$.}
	\label{fig:energy_intensity_3}
\end{figure*}

In Fig. $4$ we again consider the same parameters as in Fig. $1$, with the
exception of $N=10^{6}$, resulting in the higher values for both
$N\gamma/\omega_{0}=1$ and $\alpha=2\times10^{2}$. Here, the substantial
increased of the decay rate $\gamma$ relative to $\omega_{0}$ causes a strong
decrease in the number of superpulses, such that $\tau_{c}/\tau_{1}\approx1$.
The intensity, however, is extremely high compared to that in Fig. $1$ due to
the increases in both $N$ and $\alpha$.

\begin{figure*}[t]
	\centering
	\begin{minipage}{0.48\linewidth}
		\centering
		\includegraphics[width=\linewidth]{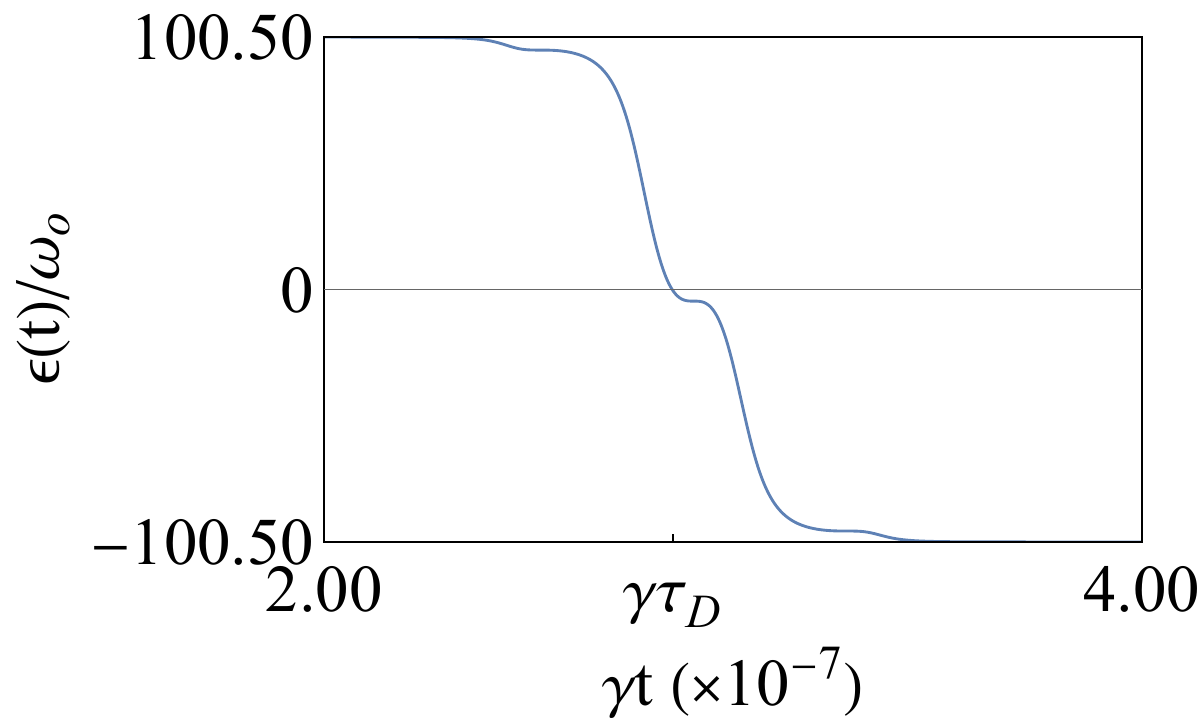}
		\\[-2pt](a)
	\end{minipage}\hfill
	\begin{minipage}{0.48\linewidth}
		\centering
		\includegraphics[width=\linewidth]{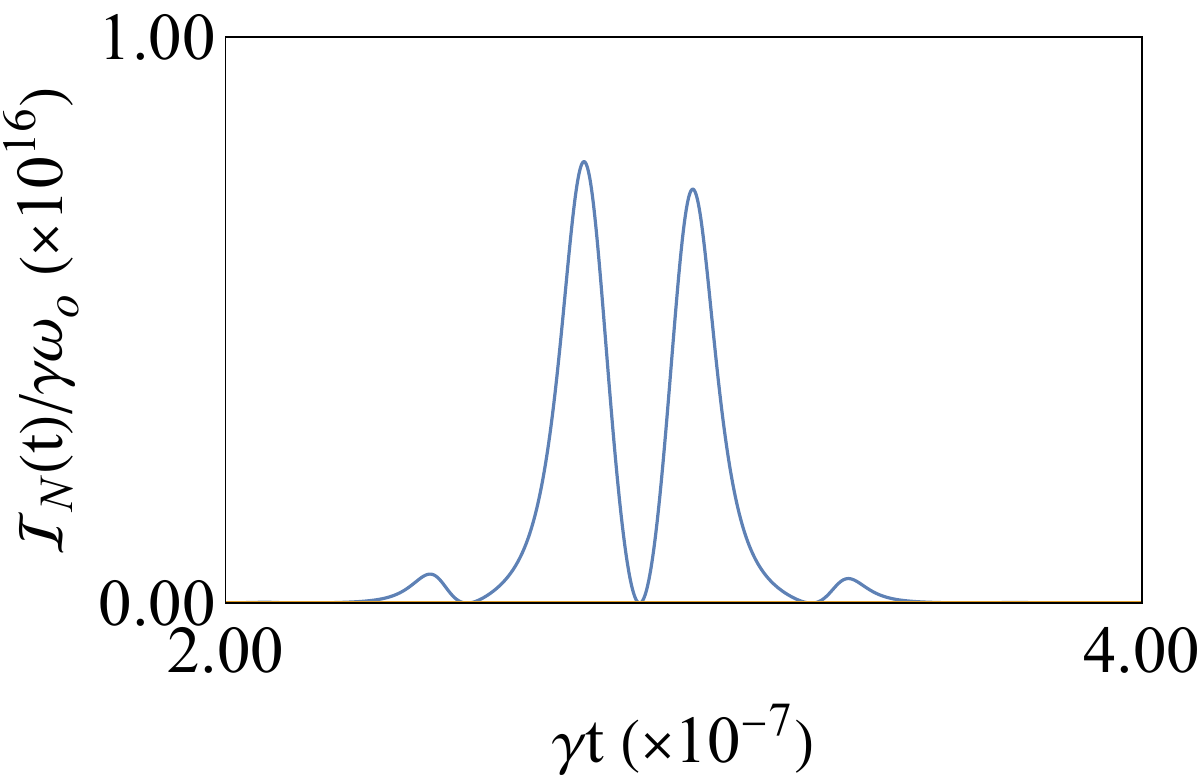}
		\\[-2pt](b)
	\end{minipage}
	\caption{Plot of the scaled (a) mean energy $\varepsilon(t)/\omega_{0}$ and
		(b) intensity $\mathcal{I}(t)/(\gamma\omega_{0})$ against $\gamma t$, for
		$N=10^{6}$, $\omega_{0}=10^{6}\gamma$, and $g=10^{2}\gamma$.}
	\label{fig:energy_intensity_4}
\end{figure*}

To illustrate the occurrence of a single superradiant pulse, in Fig. $5$ we
consider the same parameters as in Fig. $1$, with the exception of $N=10^{7}$,
resulting in the higher values for both $N\gamma/\omega_{0}=10$ and
$\alpha=2\times10^{3}$. Indeed, we observe in Fig. $5(b)$ a deformed pulse,
with the intensity increasing slowly until close to the delay time, when there
is an abrupt increase and decrease in the rate of energy change.

\begin{figure*}[t]
	\centering
	\begin{minipage}{0.48\linewidth}
		\centering
		\includegraphics[width=\linewidth]{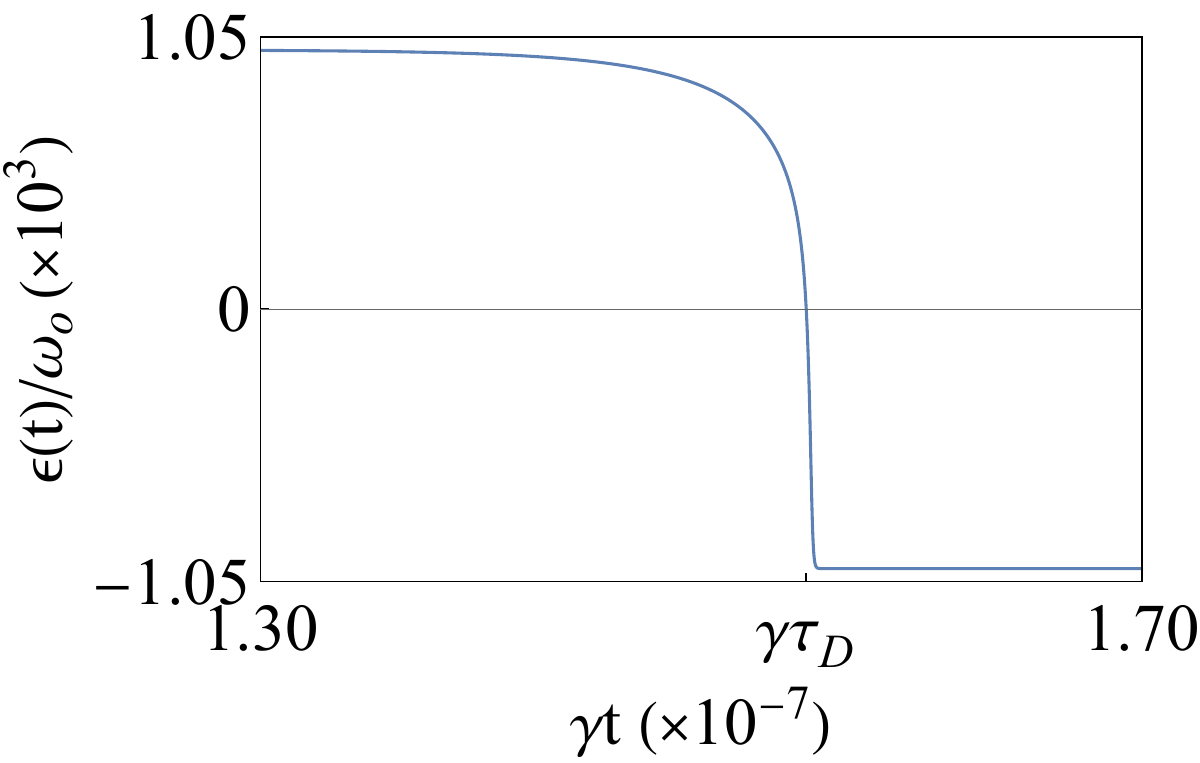}
		\\[-2pt](a)
	\end{minipage}\hfill
	\begin{minipage}{0.48\linewidth}
		\centering
		\includegraphics[width=\linewidth]{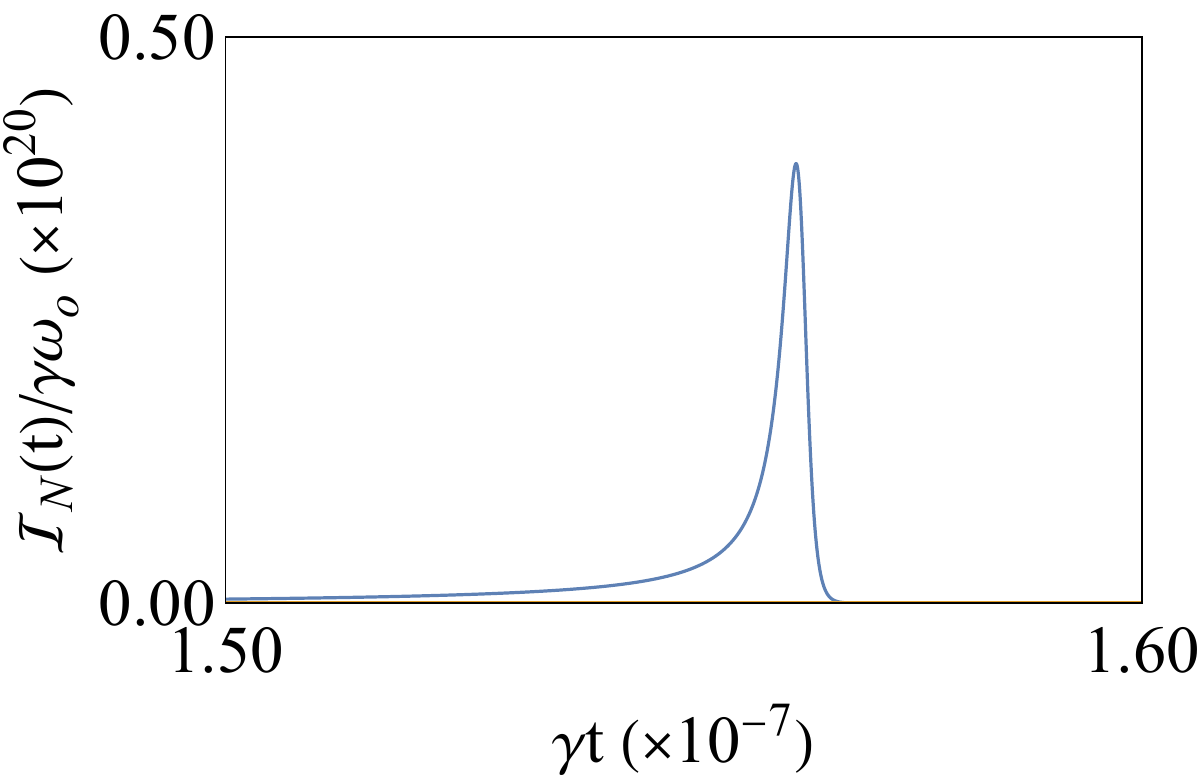}
		\\[-2pt](b)
	\end{minipage}
	\caption{Plot of the scaled (a) mean energy $\varepsilon(t)/\omega_{0}$ and
		(b) intensity $\mathcal{I}(t)/(\gamma\omega_{0})$ against $\gamma t$, for
		$N=10^{7}$, $\omega_{0}=10^{6}\gamma$, and $g=10^{2}\gamma$.}
	\label{fig:energy_intensity_5}
\end{figure*}

\subsection{Origin of Superpulses}

Now, we consider the case of a moderately dense atomic samples ($g=0$)
strongly coupled to the reservoir. In Fig. $6$ we consider $N=10^{4}$ and
$\omega_{0}=10^{6}\gamma$ to plot the scaled intensity $\mathcal{I}%
(t)/\gamma\omega_{0}$ against $\gamma t$. We verify that even in the absence
of dipolar interaction between atoms, the superpulse comb occurs, showing that
it results from the strong sample-reservoir coupling. Indeed, in the strong
coupling limit, additional terms appear in the Liouvillian (\ref{6}), which
are absent from the Liouvillian (\ref{17}) for the weak coupling limit. In
fact, considering the same parameters as in Fig. $6$, but in the weak
sample-reservoir coupling, in Fig. $7$ we verify that the scaled intensity
$\mathcal{I}(t)/\gamma\omega_{0}$ presents the usual Dicke superradiant pulse.
Therefore, as antecipated above, the sample-reservoir coupling strength is as
important in constructing the superradiant emission from dense samples as the
dipolar coupling between atomos. Finally, in Fig. $8$ we consider $N=10^{4}$
and $\omega_{0}=10^{3}\gamma$ to show that the high value $N\gamma/\omega
_{0}=10$ again causes the deformation of the superradiant pulse, as in Fig.
$5$. All equations from (\ref{24}) to (\ref{28}) remain evidently valid for
Figs. $6$ to $8$.

\begin{figure}[t]
        \centering
        \includegraphics[width=\textwidth,height=50mm,keepaspectratio]{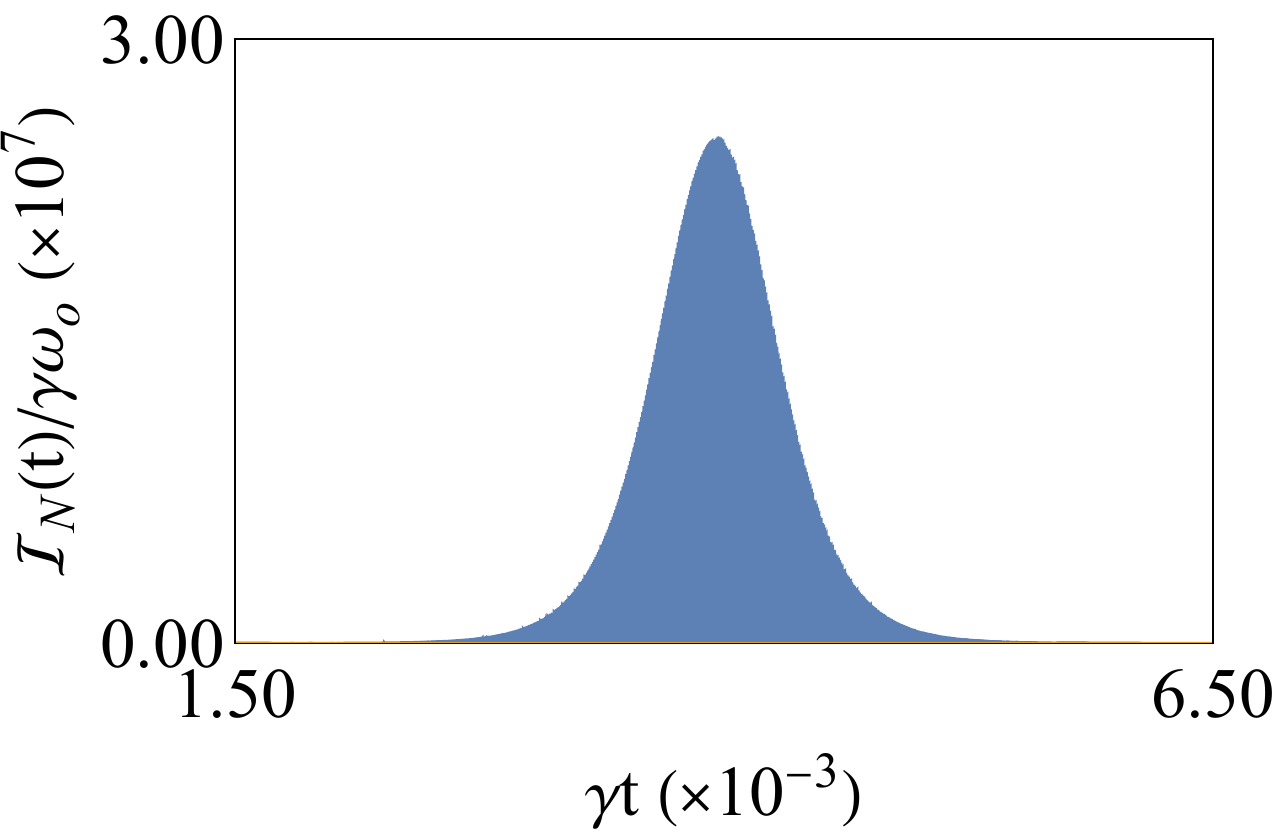}
        \caption{Plot of the scaled intensity $\mathcal{I}(t)/\gamma\omega_{0}$ against
$\gamma t$, for $N=10^{4}$, $\omega_{0}=10^{6}\gamma$ and $g=0$.}
\end{figure}

\begin{figure}[t]
        \centering
        \includegraphics[width=\textwidth,height=50mm,keepaspectratio]{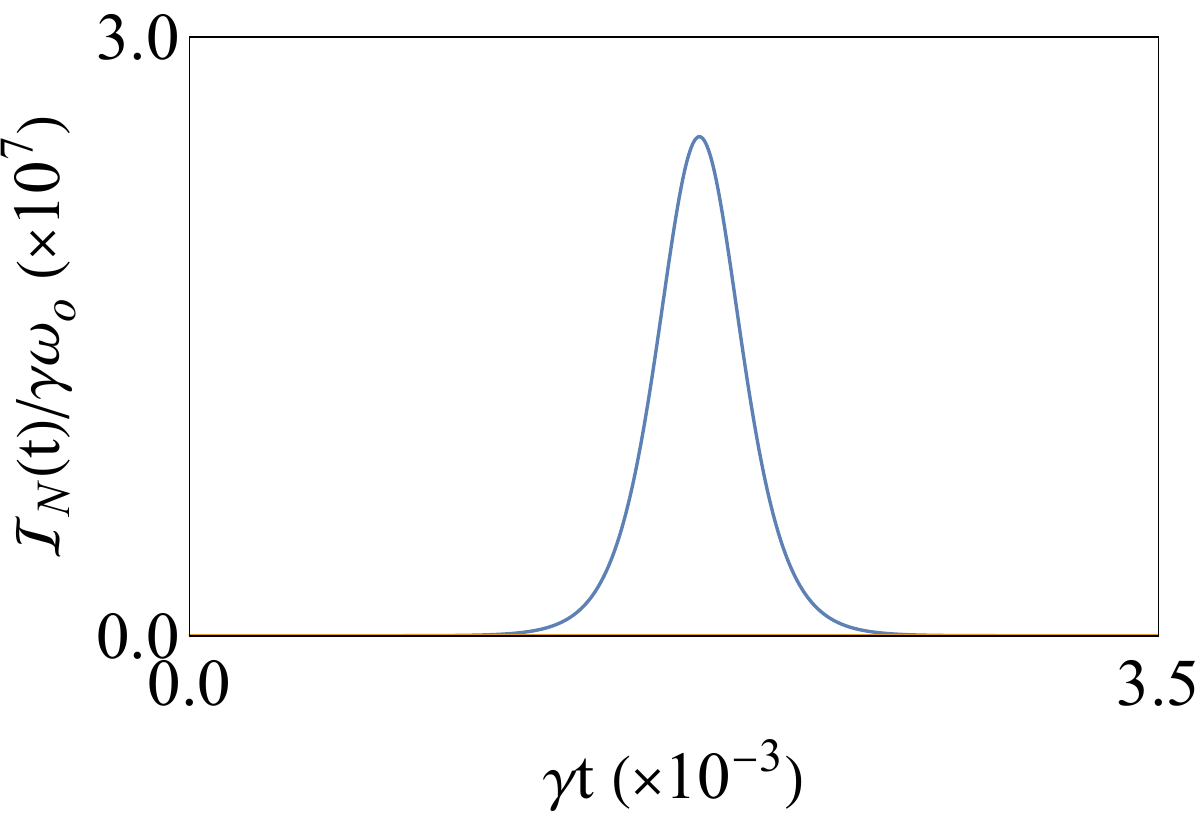}
        \caption{Plot of the scaled intensity $\mathcal{I}(t)/\gamma\omega_{0}$ against
$\gamma t$, for $N=10^{4}$, $\omega_{0}=10^{6}\gamma$ and $g=0$, considering
weak sample-reservoir coupling}
\end{figure}

\begin{figure}[t]
        \centering
        \includegraphics[width=\textwidth,height=50mm,keepaspectratio]{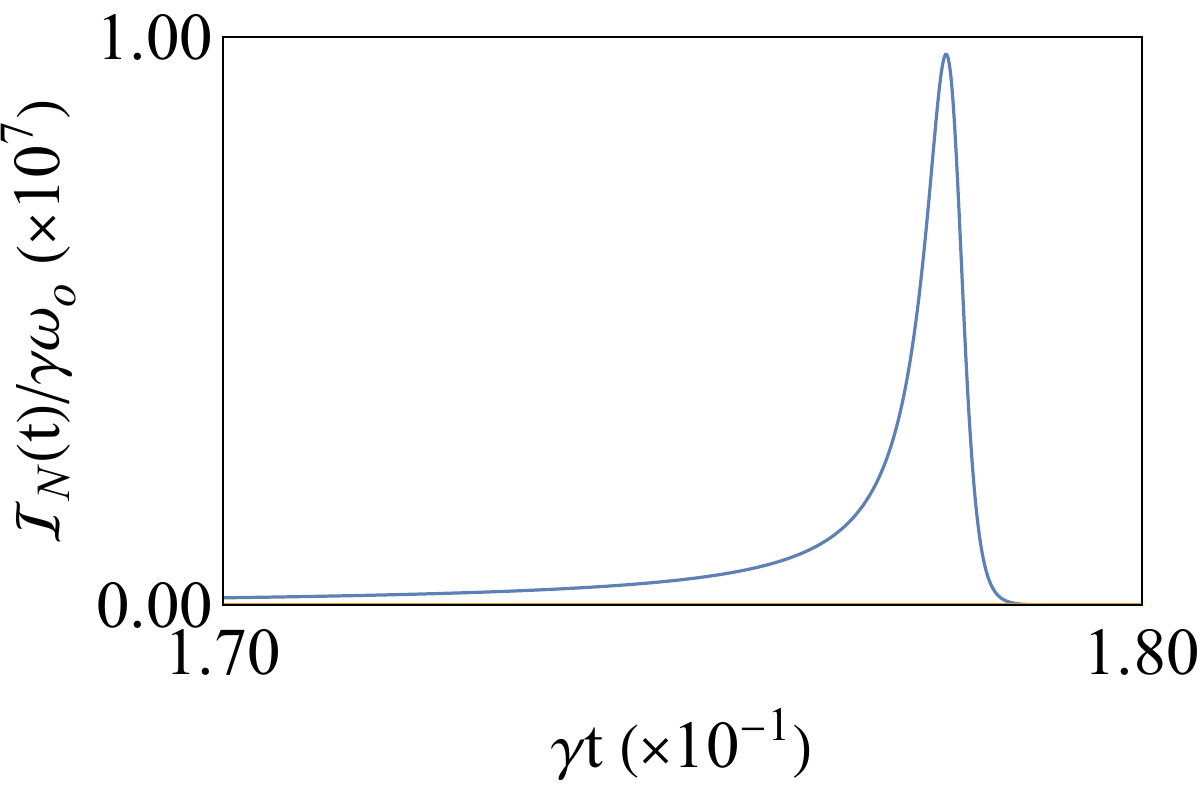}
        \caption{Plot of the scaled intensity $\mathcal{I}(t)/\gamma\omega_{0}$ against
$\gamma t$, for $N=10^{4}$, $\omega_{0}=10^{3}\gamma$ and $g=0$.}
\end{figure}

\section{Conclusions}

We present here an approach to the problem of superradiance in dense samples
coupled strongly or weakly to the reservoir. Our approach considers the
Holstein-Primakoff and mean-field approximations for the derivation of the
master equation of the problem. We verify that in the limit of moderately
dense samples, with the absence of dipolar coupling between atoms, we recover
the Dicke superradiance master equation, a good validation for the
effectiveness of our method.

We then analyze the superradiant decay from dense samples, first verifying
analytically that in the case of weak sample-reservoir coupling, the emitted
field intensifies the distinctive features of Dicke superradiance: Increasing
the intensity, from $\mathcal{I}(t)/\gamma\omega_{0}\sim\left(  N/2\right)
^{2}$ to $\left[  \left(  1+\alpha\right)  N\right]  ^{2}/4$, and decreasing
the characteristic time (from $\gamma\tau_{c}\sim1/N$ to $\gamma\tau_{c}%
\sim1/\left(  1+\alpha\right)  N$) of the Dicke pulse. In the case of strong
sample-reservoir coupling, we saw the emergence of a comb of superpulses
enveloped by the already reduced characteristic emission time $1/\left(
1+\alpha\right)  N$ of the case of weak sample-reservoir coupling. The
superpulses that make up the envelope have, in turn, even shorter
characteristic times, $\tau_{1}\sim\left[  (1+\alpha)\omega_{0}\right]  ^{-1}%
$, revealing a new scenario in radiation emission. The emergence of
superpulses is due to the Liouvillian that defines the strong sample-reservoir
coupling; however, the atomic decay rate cannot be exceedingly high, otherwise
the combo is reduced to a single superpulse.

It is worth noting that dipole-dipole-like interactions between atoms can be
engineered for a moderately dense sample interacting dispersively with a
cavity mode, as shown in Ref. \cite{zheng}. This possibility of simulating a
dense sample through a moderately dense one, makes the experimental
verification of the present proposal much more attractive.

Finally, we mention that superradiant emission has potential technological
applications, from lasers \cite{Laser, Laser2} to quantum sensors \cite{QS}, metrology
\cite{M} and cryptography \cite{C}. We believe that the distionctive features
of the  results presented here for the emission of radiation from dense
samples may motivate new investigations, both conceptual and practical, of
superradiant light.

\section*{Acknowledgements}
The authors would like to thank CAPES and FAPESP for support.

\bibliographystyle{apsrev4-2}
\bibliography{manuscript.bib}
\end{document}